\documentclass[aps,prl,onecolumn,amsmath,amssymb,superscriptaddress]{revtex4}
\usepackage{graphicx}
\usepackage{amssymb}
\usepackage{natbib}
\usepackage{color}
\usepackage[utf8]{inputenc}
\usepackage{caption}

\begin{document}

\title{Strong local moment antiferromagnetic spin fluctuations in V-doped LiFeAs}

\author{Zhuang Xu}
\thanks{These authors made equal contributions to this paper}
\affiliation{Center for Advanced Quantum Studies and Department of Physics, Beijing Normal University, Beijing 100875, China}

\author{Guangyang Dai}
\thanks{These authors made equal contributions to this paper}
\affiliation{Beijing National Laboratory for Condensed Matter Physics, Institute of Physics, Chinese Academy of Sciences,Beijing 100190, China}

\author{Yu Li}
\affiliation{Louisiana Consortium for Neutron Scattering and Department of Physics and Astronomy, Louisiana State University, Baton Rouge, Louisiana 70803, USA}
\affiliation{Department of Physics and Astronomy, Rice University, Houston, Texas 77005, USA}

\author{Zhiping Yin}
\email{yinzhiping@bnu.edu.cn}
\affiliation{Center for Advanced Quantum Studies and Department of Physics, Beijing Normal University, Beijing 100875, China}

\author{Yan Rong}
\affiliation{Center for Advanced Quantum Studies and Department of Physics, Beijing Normal University, Beijing 100875, China}

\author{Long Tian}
\affiliation{Center for Advanced Quantum Studies and Department of Physics, Beijing Normal University, Beijing 100875, China}

\author{Panpan Liu}
\affiliation{Center for Advanced Quantum Studies and Department of Physics, Beijing Normal University, Beijing 100875, China}

\author{Hui Wang}
\affiliation{Center for Advanced Quantum Studies and Department of Physics, Beijing Normal University, Beijing 100875, China}

\author{Lingyi Xing}
\affiliation{Beijing National Laboratory for Condensed Matter Physics, Institute of Physics, Chinese Academy of Sciences,Beijing 100190, China}
\affiliation{Department of Physics and Astronomy, Louisiana State University, Baton Rouge, Louisiana 70803,USA}

\author{Yuan Wei}
\affiliation{Beijing National Laboratory for Condensed Matter Physics, Institute of Physics, Chinese Academy of Sciences,Beijing 100190, China}

\author{Ryoichi Kajimoto}
\affiliation{Neutron Science Section, Materials and Life Science Division, J-PARC Center, Tokai, Ibaraki 319-1195, Japan}

\author{Kazuhiko Ikeuchi}
\affiliation{Neutron Science and Technology Center, Comprehensive Research Organization for Science and Society (CROSS), Tokai, Ibaraki 319-1106, Japan}

\author{D. L. Abernathy}
\affiliation{Quantum Condensed Matter Division, Oak Ridge National Laboratory, Oak Ridge, Tennessee 37831, USA}

\author{Xiancheng Wang}
\affiliation{Beijing National Laboratory for Condensed Matter Physics, Institute of Physics, Chinese Academy of Sciences,Beijing 100190, China}

\author{Changqing Jin}
\affiliation{Beijing National Laboratory for Condensed Matter Physics, Institute of Physics, Chinese Academy of Sciences,Beijing 100190, China}

\author{Xingye Lu}
\affiliation{Center for Advanced Quantum Studies and Department of Physics, Beijing Normal University, Beijing 100875, China}

\author{Guotai Tan}
\email{tangt@bnu.edu.cn}
\affiliation{Center for Advanced Quantum Studies and Department of Physics, Beijing Normal University, Beijing 100875, China}

\author{Pengcheng Dai}
\email{pdai@rice.edu}
\affiliation{Department of Physics and Astronomy, Rice University, Houston, Texas 77005, USA}
\affiliation{Center for Advanced Quantum Studies and Department of Physics, Beijing Normal University, Beijing 100875, China}

\date{\today}

\begin{abstract}
We use neutron scattering to study vanadium (hole)-doped LiFe$_{1-x}$V$_x$As. In the undoped state, LiFeAs exhibits superconductivity at $T_c=18$ K and transverse
incommensurate spin excitations similar to electron overdoped iron pnictides. Upon vanadium-doping to form LiFe$_{0.955}$V$_{0.045}$, the transverse
incommensurate spin excitations in LiFeAs transform into longitudinally elongated in a similar fashion as that of potassium (hole) doped Ba$_{0.7}$K$_{0.3}$Fe$_2$As$_2$, but with dramatically enhanced magnetic scattering and elimination of superconductivity. This is different from
the suppression of the overall magnetic excitations in hole doped BaFe$_2$As$_2$ and the enhancement of superconductivity near optimal hole doping.  These results are consistent with density
function theory plus dynamic mean field theory calculations, suggesting that vanadium-doping in LiFeAs may induce an enlarged effective magnetic moment $S_{eff}$ with a spin crossover
ground state arising from the inter-orbital scattering of itinerant electrons.
\end{abstract}

\maketitle

\section{introduction}
The flexible spin states of Fe$^{2+}$ ions are interesting due to its extensive distribution, from the lowermost mantle in the earth \cite{Mantle}, to the innermost hemoglobin in our human body \cite{Hemoglobin}, from the most common ferrite magnet, to the most recent iron based superconductors \cite{stewart,Frank2009,Dai_RMP}. Ignoring the orbital angular momentum, each Fe$^{2+}$ ion with six electrons distributed in five $d$ orbitals has three possible spin states, $S=0,$ 1, or 2 [Fig. 1(a)]. In iron based superconductors, where Fe$^{2+}$ ion is surrounded by crystal electric field of As atoms, the five-fold degeneracy of the $d$ orbitals is split into a two-fold degenerate $e_g$ and a threefold degenerate $t_{2g}$ states nearby the Fermi level [Fig. 1(a)] \cite{stewart}. Although it is widely believed that Hund's rule plays a critical role in determining the electronic configuration of iron \cite{Yin2011}, the $S=2$ spin state has never been observed in metallic iron pnictides \cite{Dai_RMP,norman}.

The total moment sum rule within a Heisenberg model for system with spin $S$ requires $\left\langle m^2\right\rangle = (g\mu_B)^2S(S+1)$, where $g\approx 2$ is the Land$\rm \acute{e}$ $g$-factor, when magnetic scattering is integrated over all energy and momentum space within the Brillouin Zone (BZ) \cite{Lorenzana2005}. Since neutron scattering directly measures the energy ($E$) and momentum (${\bf Q}$) dependence of the dynamic structure factor $S({\bf Q},E)$, it is therefore possible to estimate the effective spin fluctuating moment $S_{eff}$ by integrating the magnetic spectral weight over all energies and wave vectors within the BZ. In previous inelastic neutron scattering on electron-doped BaFe$_2$As$_2$ compounds, it is established that electron-doping via Ni-substitution in BaFe$_{2-x}$Ni$_x$As$_2$ reduces slightly the total spin fluctuating moment, changing from $\left\langle m^2\right\rangle\approx 3.5$ $\mu_B^2$/Fe at $x=0$ to $\left\langle m^2\right\rangle\approx 2.7$ $\mu_B^2$/Fe at $x=0.3$ \cite{Meng2013,Luo2012,Luo2013}. These values approximately correspond to $S_{eff}\approx 1/2$ \cite{Meng2013,Liu2012}. For hole-doped Ba$_{1-x}$K$_x$Fe$_2$As$_2$, the total spin fluctuating moment decreases rapidly with increasing $x$ \cite{Meng2013}, and $S_{eff}$ reduces to $0.14$ at $x=1$ \cite{Horigane2016}. For electron-doped NaFe$_{1-x}$Co$_x$As \cite{Dai_RMP}, the total spin fluctuating moment also decreases slightly with increasing $x$ but has almost identical value as that of BaFe$_{2-x}$Ni$_x$As$_2$ \cite{Scott2016}, thus suggesting universal nature of the $S_{eff}\approx 1/2$ for two different families of iron pnictides.
{This is different from the superconducting LiFeAs, where the low-energy spin fluctuations coupled to superconductivity
are about a factor 4 smaller than that of the Co-underdoped
superconducting BaFe$_2$As$_2$ \cite{Qureshi2012,Qureshi2014}. The total fluctuating moment
of LiFeAs is $\left\langle m^2\right\rangle\approx 1.5\pm 0.3$ $\mu_B^2$/Fe, more than a factor of 2 smaller
than that of BaFe$_2$As$_2$ \cite{Wang2012,Yu2016}.}
On the other hand, the total spin fluctuating moment for iron chalcogenide compounds FeSe and FeTe are significantly larger. In the case of FeSe, the total spin fluctuating moment of $\left\langle m^2\right\rangle\approx 5.2$ $\mu_B^2$/Fe arises entirely from dynamic magnetic excitations as the system has no static ordered moment \cite{Wang2016}. For FeTe, the effective moment of Fe may change from $S_{eff}\approx 1$ at 10 K to $S_{eff}\approx 3/2$ at 300 K, suggesting entangled local magnetic moments with the itinerant electrons on warming \cite{Zaliznyak2011}.
The total moment sum rule strictly speaking is only valid for insulating local moment systems, where quantum effect and itinerant electron induced magnetism are not important \cite{Lorenzana2005}. In iron-based superconductors, itinerant electrons and hole-electron Fermi surface nesting are known to be important for magnetism and superconductivity [Fig. 1(b)] \cite{hirschfeld}. Theoretically, many physical properties of iron based superconductors can be understood within the orbital selective Mott phase (OSMP) picture, where electrons in Fe$^{2+}$ ion with different orbitals can behave differently \cite{Medici2014,yu2017,komijani2017}. In this model, hole-doped compounds
are believed to be more correlated and therefore should have large effective local moments \cite{Medici2014,yu2017,komijani2017}. From combined density function theory (DFT) and dynamic mean field theory (DMFT) calculations \cite{Yin2011},  the total fluctuating moments in the paramagnetic state are nearly invariant around 2.2 $\mu_B$/Fe in a wide variety of iron pnictide/chalcogenide families. In x-ray emission spectroscopy (XES) experiments, values of magnetic spin moments in the paramagnetic phase are about 1.3 $\mu_B$/Fe for iron pnictides including BaFe$_2$As$_2$ and Ba(Fe,Co)$_2$As$_2$ \cite{norman}. These values are considerably smaller than the predictions from DFT+DMFT calculation but larger than those from neutron scattering experiments \cite{Meng2013,Luo2012,Luo2013}, possibly due to the fast time scale of the XES measurements \cite{norman}. Therefore, to fully understand the key ingredients of magnetism in iron based superconductors, it is important to study the evolution of $S_{eff}$ in different families of hole-doped iron based superconductors in a controlled and systematic way.
\begin{figure}[!t]
\includegraphics[scale=.4]{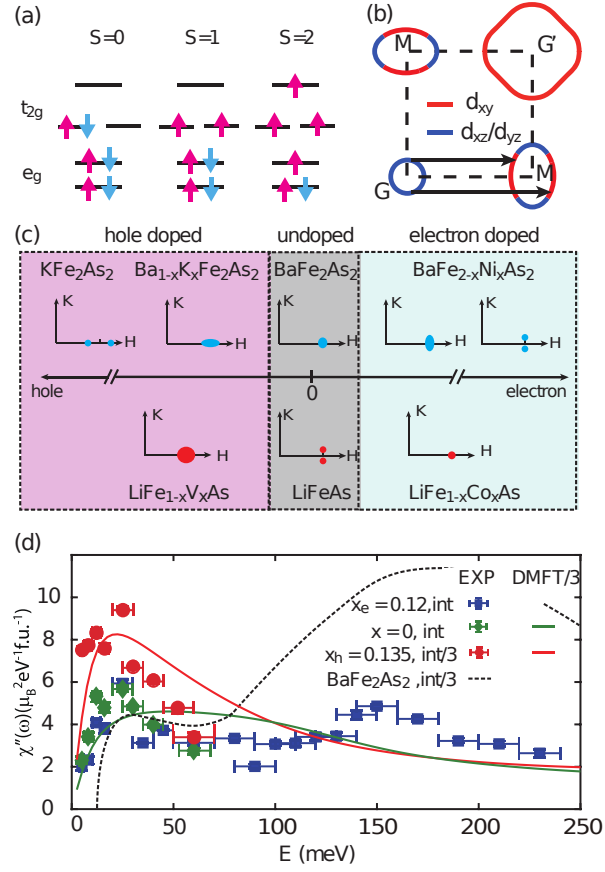}
\caption{(a) Possible spin states of Fe$^{2+}$ in $t_{2g}$ and $e_{g}$ states within a localized
moment picture. (b) Fermi surfaces of LiFe$_{0.955}$V$_{0.045}$As. Red and blue colors represent d$_{xy}$ and d$_{xz/yz}$ orbital characters, respectively. (c) Summary of the doping dependence of low energy spin
excitations in BaFe$_2$As$_2$ (Top) and LiFeAs (bottom). The shapes of the magnetic scattering are marked by blue (BaFe$_2$As$_2$) and red (LiFeAs) colored
points. It is worth noting that spin excitations in LiFe$_{0.955}$V$_{0.045}$As are
considerably diffusive in reciprocal space. (d) Theoretical and experimental local spin susceptibility in LiFeAs ($x=0$), LiFe$_{0.88}$Co$_{0.12}$As ($x_e$=0,12), LiFe$_{0.955}$V$_{0.045}$As ($x_h$=0.135) and BaFe$_2$As$_2$.}
\end{figure}

While it is easy to dope holes in BaFe$_2$As$_2$ family of iron pnictides by substituting Ba$^{2+}$ cation with K$^{1+}$, replacing Fe with Cr/Mn introduces impurities and/or disorders instead of holes \cite{Mn1,Mn2,Tucker2012}. For NaFeAs family, systematic neutron scattering and angle-resolved photoemission spectroscopy (ARPES) experiments reveal that replacing Fe with Cu dopes holes into the system and induces large ordered magnetic moment for $x>0.1$ in NaFe$_{1-x}$Cu$_x$As \cite{Song2016,Matt2016}. However, there are no inelastic neutron scattering experiments to determine the total fluctuating moment for this family of compounds. In the case of LiFeAs family, ARPES experiments indicate that doping V into Fe sites actually introduces hole carriers with a rate of 0.3 hole per V dopant and selectively enlarges the inner hole Fermi surface [Fig. 1(b)] \cite{Xing2016}, even though the doping-induced impurity and disorder effects might still play a role in the transport and susceptibility measurements with higher V-doping ratio.  Although hole-doping via K substitution in Ba$_{1-x}$K$_{x}$Fe$_2$As$_2$ induces superconductivity, less than 2\% V-doping in LiFe$_{1-x}$V$_x$As quickly suppresses the superconductivity in pure LiFeAs, even with a perfect nesting condition established between the inner hole and electron Fermi pockets at $x=0.084$
[see arrows in Fig. 1(b)] \cite{Xing2016}. The rapid suppression of superconductivity in LiFe$_{1-x}$V$_x$As has been suggested as due to magnetic impurity effect of V dopant, but there are no  neutron scattering experiments to establish the effect of V-doping to spin excitations of pure LiFeAs
\cite{Qureshi2012,Qureshi2014,Wang2012}.

\section{Results}
We report the magnetic excitations in hole-doped nonsuperconducting LiFe$_{0.955}$V$_{0.045}$As, which has nearly nested hole-electron Fermi surfaces with $d_{xz/yz}$ and $d_{xy}$ orbital characters, respectively, and exhibits non-Fermi-liquid behavior \cite{Xing2016}. We observe enhanced commensurate magnetic fluctuations at antiferromagnetic (AF) wave vectors, suggesting that the non-Fermi-liquid behavior is intimately associated with magnetic fluctuations. The broadness and diffuseness of the spin excitations are also distinct from the sharp commensurate spin fluctuations in electron-doped LiFe$_{0.88}$Co$_{0.12}$As \cite{Yu2016}, probably reflecting the local nature of quantum spin fluctuations \cite{Ong2012}. More importantly, we find that the intensities of spin fluctuations are strongly enhanced in the energy range of our measurements with a larger $S_{eff}$, in sharp contrast to the reduced effective moment in hole-doped Ba$_{1-x}$K$_x$Fe$_2$As$_2$ \cite{Meng2013,Horigane2016}. Considering the different superconducting behavior in Ba$_{1-x}$K$_x$Fe$_2$As$_2$ and LiFe$_{1-x}$V$_x$As, we argue that the inter-orbital scattering processes between $d_{xz/yz}$ and $d_{xy}$ orbitals increase the fluctuating moment but are detrimental to superconductivity \cite{JHZhang2000}.  These results are consistent with the notion that intra-orbital scattering processes between hole and electron pockets, particularly those involving $d_{yz}$-$d_{yz}$ orbital characters, are good for superconductivity in different classes of iron-based superconductors \cite{TChen2019,Pfau,Pfau2019,Watson2019,LTian2019}.

\begin{figure}[!t]
\includegraphics[scale=.35]{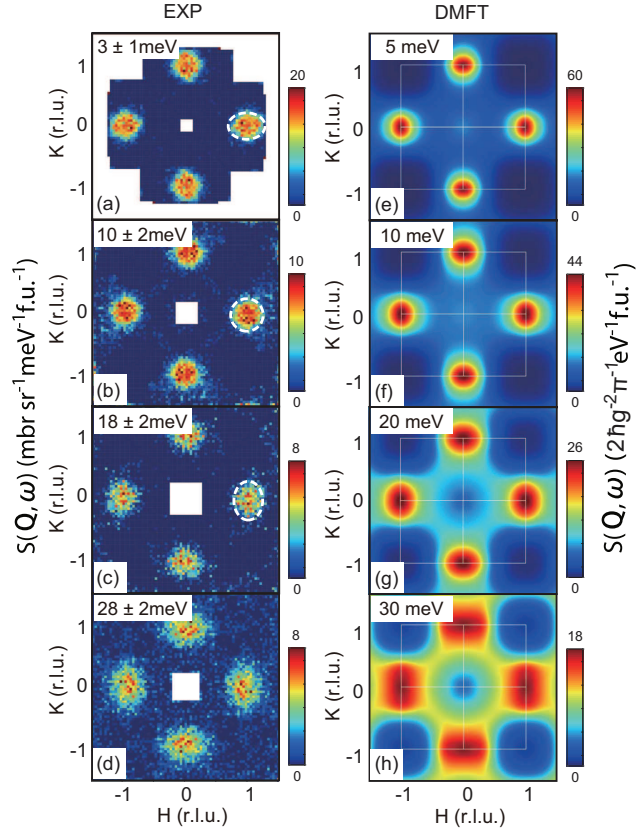}
\caption{(a-d) 2D images of measured dynamic spin susceptibility of LiFe$_{0.955}$V$_{0.045}$As in the $[H,K]$
plane at $E = 3 \pm 1$, $10\pm 2$, $18\pm 2$ and $28\pm 2$ meV, respectively. The elongation of spin excitations along [H,0,0] can be observed in both experimental (dashed circle) and theoretical (green and red area) results.  The data were collected at $T=5$ K and folded to improve statistics with radially symmetric backgrounds subtracted. (e-h) The corresponding results from DFT+DMFT calculations. {The color bars indicate scattering intensity in absolute units from experiments and theory.}}
\end{figure}

\begin{figure}[!t]
\includegraphics[scale=.45]{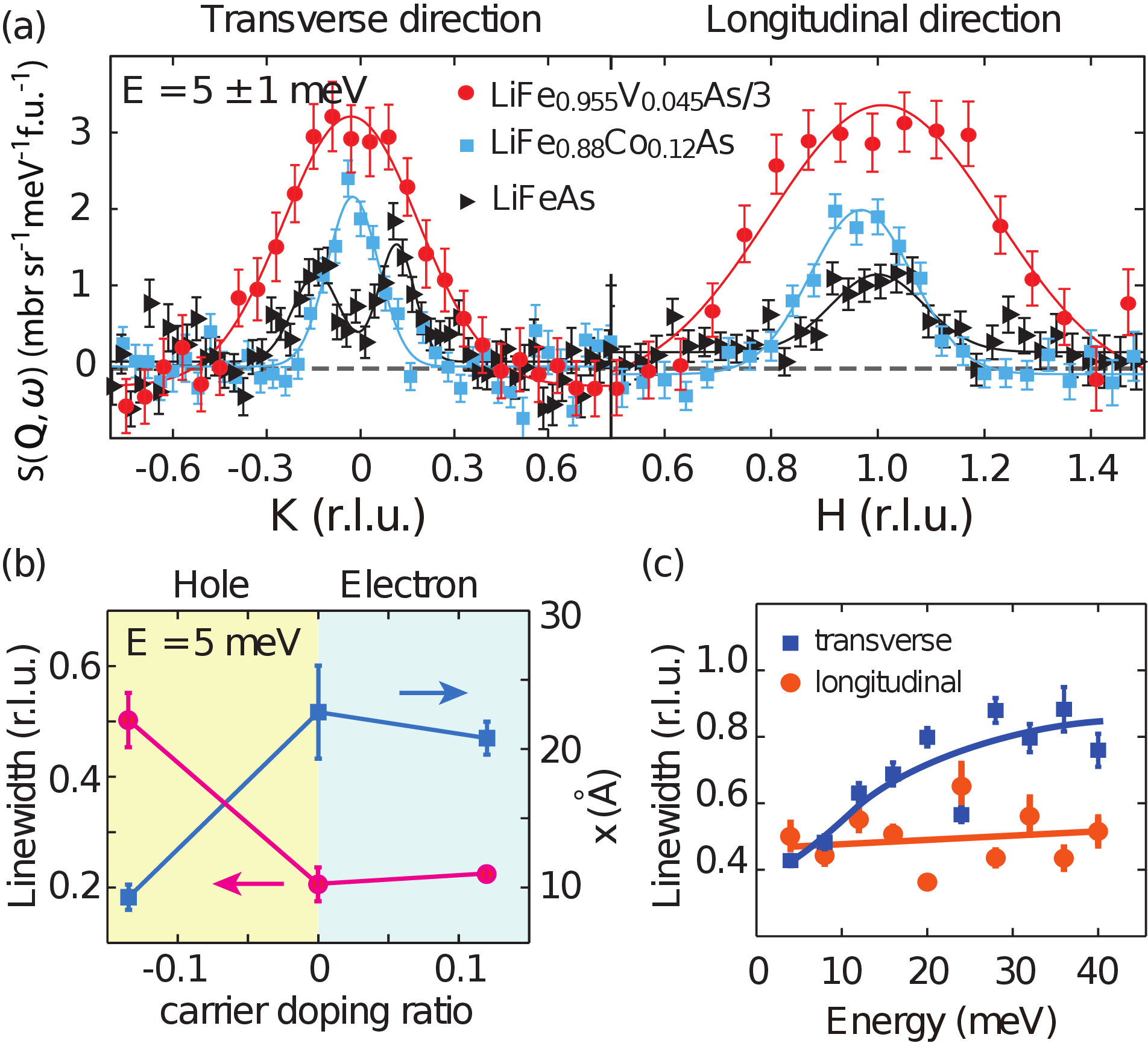}
\caption{(a) Constant-energy cuts of spin excitations along the transverse $[1,K]$ (left) and longitudinal $[H,0]$ (right) directions for LiFeAs, LiFe$_{0.88}$Co$_{0.12}$As, and LiFe$_{0.955}$V$_{0.045}As$ at $E = 5\pm 1$ meV, respectively. The solid lines are corresponding single- or two-Gaussian fits. The intensity is normalized by relative phonon intensity to account for spectrometer differences \cite{Scott2016}. (b) Line width and correlation length of the low energy spin fluctuations along {the longitudinal $[H,0]$}  direction in LiFeAs and hole/electron doped compounds. The doping ratio is a theoretical value assuming each V dopant contributes 3 holes. (c) Energy dependence of the line widths of the low energy spin fluctuations in LiFe$_{0.955}$V$_{0.045}As$ along the transverse and longitudinal directions, respectively. }
\end{figure}

We first plot in Figure 2 the two-dimensional (2D) images of spin excitations in LiFe$_{0.955}$V$_{0.045}$As at different energies and their comparison with the DFT+DMFT calculations for 10\% hole doping (about LiFe$_{0.967}$V$_{0.033}$As). At energy transfer $E = 3\pm 1$ meV, spin excitations occur at commensurate $Q_{AF} = (1,0)$ and (0,1) positions [Figs. 1(c), 2(a)], different from the transverse incommensurate spin fluctuations in LiFeAs \cite{Qureshi2012,Wang2012,Qureshi2014}.
Instead, spin excitations display a small elongation along the longitudinal direction, mimicking the low-energy spin excitations of hole-doped Ba$_{1-x}$K$_{x}$Fe$_2$As$_2$ \cite{CZhang2011}.
As the energy transfer increases from $E = 3$ to 18 meV, the elongation of the elliptical spin fluctuations changes from the longitudinal to transverse direction, similar to that in Ba$_{0.67}$K$_{0.33}$Fe$_2$As$_2$ \cite{Meng2013}, reflecting the hole-electron Fermi surface nesting contribution from itinerant electrons \cite{CZhang2011,JHZhang2000}. The wave vector dependent spectra at the corresponding energies are confirmed by the DFT+DMFT calculations in the 10\% hole-doped LiFeAs [Figs. 2(e-h)]. These results are significantly different from Mn-substituted BaFe$_2$As$_2$, where Mn-doping induces diffusive magnetic scattering at the checkerboard wave vector $(1,1)$ \cite{Tucker2012}. The broad and diffusive magnetic scattering at (1,0)/(0,1) may be attributed to the short range spin fluctuations from localized moment and/or not perfect nesting condition between hole-electron Fermi surfaces.

To compare the magnitude of spin fluctuations in LiFeAs, LiFe$_{0.88}$Co$_{0.12}$As, and LiFe$_{0.955}$V$_{0.045}$As, we calculate in Fig. 1(d) the energy dependence of
the local dynamic spin susceptibility $\chi^{\prime\prime}(E)$ for these materials obtained by integrating wave vector dependent dynamic spin susceptibility $\chi^{\prime\prime}({\bf Q},E)$ over the entire BZ \cite{Dai_RMP}. Compared with those of LiFeAs and LiFe$_{0.88}$Co$_{0.12}$As \cite{Yu2016},
$\chi^{\prime\prime}(E)$ in LiFe$_{0.955}$V$_{0.045}$As is clearly enhanced below about 80 meV, although we do not have data above 80 meV. Such an enhancement is also captured in the DFT+DMFT calculations as shown in red and blue solid lines of Fig. 1(d). The local dynamic spin susceptibility of LiFe$_{0.955}$V$_{0.045}$As in the measured energy range is somewhat
larger than that of BaFe$_2$As$_2$ \cite{Harriger,xylu18}. {For comparison,
magnetic scattering of LiFeAs is considerably smaller than that of the Co-doped
superconducting Ba(Fe$_{0.92}$Co$_{0.08}$)$_2$As$_2$ \cite{Qureshi2012,Qureshi2014}.}

The differences between LiFeAs, LiFe$_{0.88}$Co$_{0.12}$As, and LiFe$_{0.955}$V$_{0.045}$As can be further confirmed in Fig. 3(a) in which we plot constant-energy cuts of spin excitations along the transverse and longitudinal directions. In all cases, we obtain the magnitude of magnetic scattering by normalizing it to phonon intensities \cite{Scott2016}, making it possible to compare magnetic scattering of different materials measured on different instruments. It is clear that the magnetic scattering in LiFe$_{0.955}$V$_{0.045}$As is both wider and higher than those in LiFeAs and LiFe$_{0.88}$Co$_{0.12}$As at $E=5$ meV, resulting in a large increase in $\chi^{\prime\prime}(E)$ [Fig. 1(d)].
This can not be explained by the extra Vanadium magnetic impurities since the 3$d^3$ electronic configuration of V$^{2+}$ ions only has $S = 3/2$ which is smaller than
the expected $S =2$ for 3$d^6$ of Fe$^{2+}$ in iron based superconductors. Therefore, the enhanced magnetic scattering has to be associated with the introduction of hole carriers which effectively modifies the spin configuration of Fe$^{2+}$ ions.

In Fig. 3(b), we show the line width of spin excitations and the corresponding dynamic spin-spin correlation length for LiFeAs, LiFe$_{0.88}$Co$_{0.12}$As, and LiFe$_{0.955}$V$_{0.045}$As. The sudden drop of spin correlation length in LiFe$_{0.955}$V$_{0.045}$As suggests that short-range spin correlations of local moments may become important.
 Figure 3(c) shows the line widths of spin excitations along both the transverse and longitudinal directions at different energies in LiFe$_{0.955}$V$_{0.045}$As. The crossing around 10 meV is consistent with the isotropic spin excitations in Fig. 2(b). Therefore, LiFe$_{1-x}$V$_x$As is distinct from Cr/Mn doped BaFe$_2$As$_2$ in which magnetic impurities induce N$\rm \acute{e}$el-type diffusive AF spin fluctuations at low concentrations \cite{Mn1,Mn2,Tucker2012}.

\begin{figure}[!t]
\includegraphics[scale=.45]{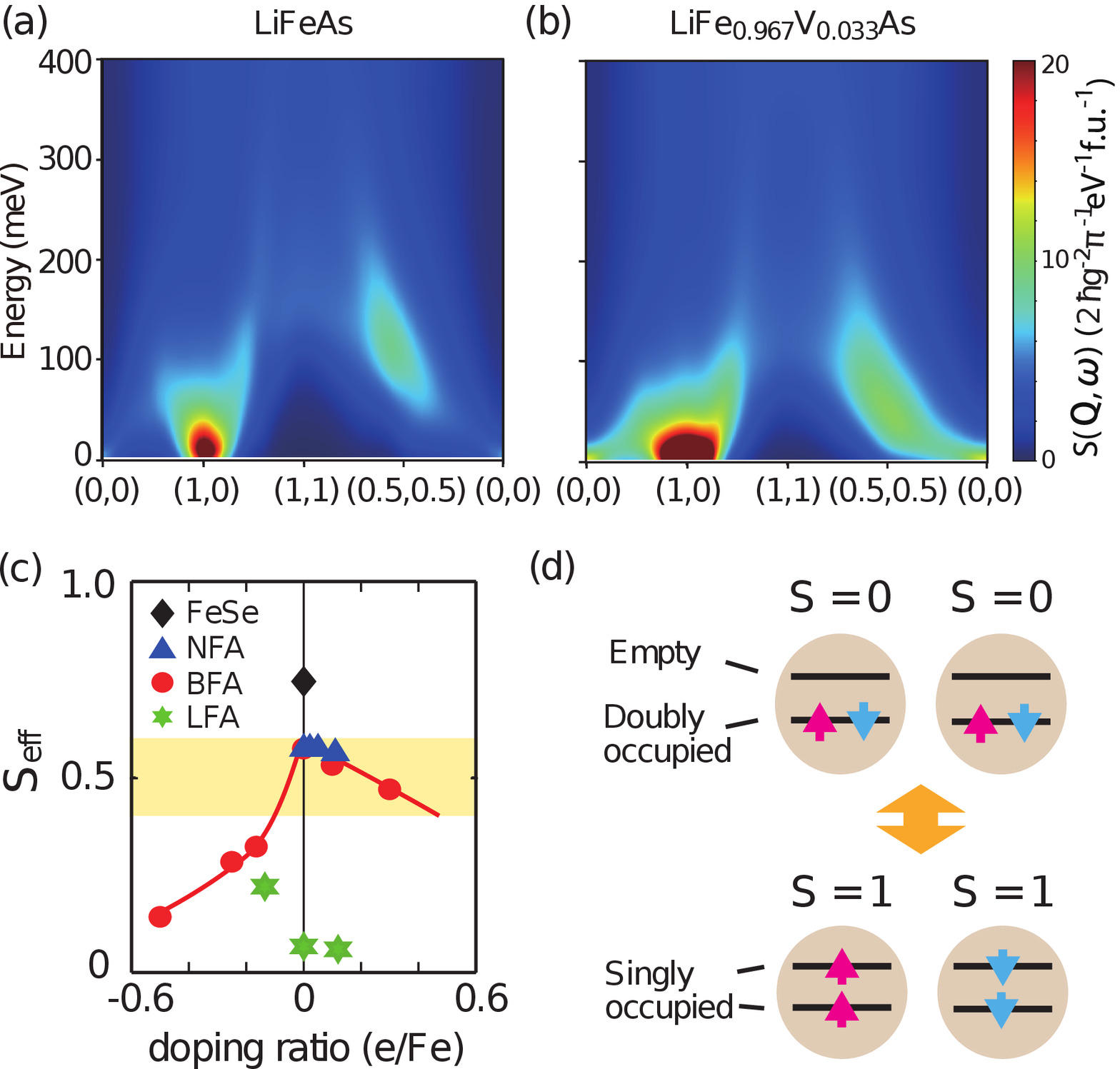}
\caption{(a,b) The calculated $S({\bf Q},E)$, obtained by using computed
dynamic spin susceptibility $\chi^{\prime\prime}({\bf Q},E)$ and
{$\chi^{\prime\prime}({\bf Q},E)=g^2\mu_B^2{\pi\over\hbar}(1-\exp(-\hbar\omega/k_BT))S({\bf Q},E)$}, of LiFeAs and 10$\%$ hole doped LiFe$_{0.967}$V$_{0.033}$As by DFT+DMFT \cite{Dai_RMP}. They are plotted in the same color scale for direct comparison.
The factor 2 in the units comes from a factor of 2 in the formula
that we didn’t multiply when we wrote the code to compute $\chi^{\prime\prime}$.
The color bar indicates scattering intensity in absolute units from the DFT+DMFT calculation.
(c) The effective spin $S_{eff}$ in various iron pnictide/calcogenides superconductors obtained from neutron scattering experiments \cite{Wang2016,Meng2011,Horigane2016,Yu2016,Scott2016,Luo2013}. NFA,BFA and LFA represent NaFeAs, BaFe$_2$As$_2$ and LiFeAs families, respectively.
The energy integration range of $\chi^{\prime\prime}(E)$ for LiFeAs family of materials is from 0 to about 60 meV, and thus does not represent the total local dynamic susceptibility.
(d) Schematics of dynamic mixing of spin states with $S=0$ and 1. }
\end{figure}

Figures 4(a) and 4(b) compare the DFT+DMFT calculated wave vector and energy dependence of $\chi^{\prime\prime}({\bf Q},E)$ for LiFeAs and LiFe$_{0.967}$V$_{0.033}$As, respectively. Consistent with Fig. 1(d), we find that LiFe$_{0.967}$V$_{0.033}$As has larger magnetic spectral weight, especially at low energies.  In Fig. 4(c), we show the estimated effective spin $S_{eff}$ for different iron based superconductors. It is clear that all values are distributed between $S=0$ and 1 with most of them around $S_{eff} = 1/2$. The discrepancy of hole doping dependence of the magnetic excitations between BaFe$_2$As$_2$ and LiFeAs families may arise mostly from the fact that we have only integrated the energy range of $\chi^{\prime\prime}(E)$ from 0 to about 60 meV where we have solid data for V-doped and pure LiFeAs. Although these numbers cannot be compared with absolute values of $S_{eff}$ determined for BaFe$_2$As$_2$ and FeSe, the trend of increasing total magnetic susceptibility with hole-doping in V-doped LiFeAs is unmistakable.
Therefore, it is interesting to understand the microscopic origin of magnetic excitations in V-doped LiFeAs.

\section{Discussion}

Theoretically, the reduction of effective spins in iron pnictides/calcogenides can be understood in either an itinerant or a localized moment picture. The former considers Fe$^{2+}$ ions embedded in a Fermi sea of conduction electrons \cite{Rev_hund,Aron2015} and the lower spin states can be achieved through a multiple-stage Kondo screening process \cite{Stadler2015,Ong2012}, where the system is in a quantum critical incoherent metallic state down to 0 K. In the local moment scenario, due to the sensitivity of spin states of Fe$^{2+}$ to the ionic radius, the effective magnetic moment in iron based material can be tuned by controlling the strength of crystal electric field in the local FeAs$_4$ tetrahedral structure \cite{Chaloupka2013,SCO}. This picture can successfully explain some phenomena such as temperature-dependent magnetic moments \cite{Gretarsson2013,Klingeler2010} and coherent-incoherent crossover \cite{Ong2012,Miao2016,Run2017,Hardy2013,Wiecki2018}.
However, it is still difficult to understand the large variety of the fluctuated moments in iron based superconductor by assuming a reduced spin state with $S = 1$ or 0. To achieve an intermediate crossover between spin state $S =1$ and $S = 0$, a dynamic ``spin mixing'' process was
proposed \cite{Chaloupka2013,SCO} as schematically shown in Fig. 4(d). When two Fe$^{2+}$ ions exchange their electrons, their spin states would effectively change from $S$ = 0 to $S$ =1 and vice versa. This results in a continuous evolution of the effective spin state with $0<S_{eff}<1$. In this scenario, the probability of each $S=0$ or 1 state is closely related to both the local structure and conduction electrons \cite{Chaloupka2013}. We note that this spin crossover scenario is a localized model and may not completely reflect the itinerant nature of iron based superconductors. Further theoretical and experimental evidences are needed to clarify the situation.

In insulating copper oxides, local moments are locked into the $S=1/2$ state by the elimination of doubly occupied states with a large charge gap \cite{Lee2006}. Doping charge carriers introduce doubly occupied states and lead to the destruction of local moments, resulting the suppression of the neutron scattering spectral weight in the energy range of typical neutron scattering experiments \cite{Lorenzana2005}. Similarly, in iron pnictides, the effective moment on each Fe$^{2+}$ ion is also affected by the population of doubly occupied states or spin singlets due to Kondo screening. However, this can not be simply attributed to the introduction of charge carriers since the doping dependence of $S_{eff}$ in Ba$_{1-x}$K$_x$Fe$_2$As$_2$ and LiFe$_{1-x}$V$_x$As is rather different. Such a difference may be rooted in the different sites of the dopants. Since Vanadium is directly introduced into the FeAs layer while K$^+$ is not, the Fe-As bond length is affected by V-doping \cite{Xing2016} and a high spin state is favored in V-doped LiFeAs. With increasing V-doping, the nesting condition improves between the inner hole Fermi surface with $d_{xz/yz}$ orbital character and the inner electron pocket mainly derived from $d_{xy}$ orbital character \cite{Xing2016}.  Therefore, the enhanced inter-orbital scattering between $d_{xz/yz}$ and $d_{xy}$ orbitals, which suppresses superconductivity \cite{JHZhang2000}, favors the ``spin mixing'' process [Fig. 4(d)] and enhances fluctuated moments. This is different from the intra-orbital
$d_{yz}$-$d_{yz}$ hole-electron Fermi surface nesting scattering
which favors superconductivity and induces
a neutron spin resonance only at the AF wave vector
in detwinned FeSe \cite{TChen2019} and underdoped iron pnictides \cite{LTian2019,WYWang}.

To conclude, we performed time-of-flight neutron scattering measurements on LiFe$_{0.955}$V$_{0.045}$As and did DFT+DMFT calculation on the corresponding 10$\%$ hole-doped LiFeAs system. We found that the low energy spin fluctuations are influenced by itinerant electrons following the same doping dependence of spin excitations in BaFe$_2$As$_2$. More importantly, both the experimental and theoretical results suggest that magnetic excitations are dramatically enhanced by doping holes with Vanadium. We discuss the possible origin of such enhancement and argue that the inter-orbital scattering between $d_{xz/yz}$ and $d_{xy}$ orbitals are bad for superconductivity and may dynamically mix the spin states of Fe$^{2+}$ ions.

\section{Methods}

{\bf Sample preparation}
We grew single crystals of LiFe$_{0.955}$V$_{0.045}$As by self-flux method \cite{Xing2014}. 7.2 grams of single crystals were used in our time-of-flight neutron scattering experiments, of which 1.8 grams were synthesized by using isotope $^7$Li to reduce neutron absorption.
Samples were wrapped by Aluminum foil with Hydrogen-free glue in the glove box filled with Argon gas since the samples are air- and humidity-sensitive. All samples were co-aligned and glued
on five $40\times 40$ mm$^2$ Aluminum plates as showed in Fig. 5(b). The $[1,0,0]$ direction is along the horizontal direction and $[1,1,0]$ along the diagonal [Fig. 5(a)].

{\bf Neutron Scattering experiments}
Our inelastic neutron scattering measurements on LiFe$_{0.955}$V$_{0.045}$As were carried out on the neutron time-of-flight Four-Dimensional (4D) Space Access Neutron Spectrometer (4SEASONS) at Materials and Life Science Experimental Facility (MLF), Japan Proton Accelerator Research Complex (J-PARC) \cite{Kajimoto}. We define the momentum transfer $Q$ in 3D reciprocal space in \AA$^{-1}$ as $\textbf{Q}=H\textbf{a}^\ast+K\textbf{b}^\ast+L\textbf{c}^\ast$, where $H$, $K$, and $L$ are Miller indices and ${\bf a}^\ast=\hat{{\bf a}}2\pi/a$, ${\bf b}^\ast=\hat{{\bf b}}2\pi/b$, ${\bf c}^\ast=\hat{{\bf c}}2\pi/c$ with  $a= b\approx 5.316$ \AA, and $c=6.315$ \AA\  \cite{Wang2012,Yu2016}. Samples are co-aligned in the $[H,0,L]$ scattering plane with mosaic less than 5$^\circ$ and incident beam ($E_i= 16,39,75$ and 200 meV) parallel to the $c$-axis of the crystals.
In principle, absolute magnetic neutron scattering intensity from LiFe$_{0.955}$V$_{0.045}$As can be normalized by comparing the scattering with a vanadium standard \cite{Dai_RMP}. However, LiFe$_{0.955}$V$_{0.045}$As has considerable unknown amount of vanadium-containing flux from the growth, rendering this method unreliable. To obtain the absolute intensity, we therefore use phonon normalization method as discussed below.

{\bf Background subtraction and data analysis}
The background subtraction processes were shown in Fig. 6 \cite{Scott2016}. The raw magnetic scattering appears in a fourfold symmetry at wave vectors $[1,0]$, $[0,1]$, $[-1,0]$, and $[0,-1]$ [Fig. 6(a)] due to the tetragonal symmetry of the lattice structure. We symmetrized the scattering spectra in order to enhance the magnetic signal and then remove the background [Fig. 6(b)]. Background was obtained by masking the signal within the white squares [Fig. 6(c)] and assumed to be radially symmetric. A polynomial function of $\left|{\bf Q}\right|$ was used to fit the background intensity [Fig. 6(e)], and then was subtracted from the raw data. The results were shown in Fig. 6(d). Figures 6(f) and 6(g) are cuts along the $[H,0]$
and $[1,K]$ directions, respectively. We note that there are still residual background intensities which likely come from phonon scattering and affect the calculation of local susceptibility. Different treatments to these intensities might result in the differences of the final estimated local susceptibility. Therefore, we recalculate the local susceptibility of pure LiFeAs and LiFe$_{0.88}$Co$_{0.12}$As from the previous data with the same method we used for V-doped LiFeAs. In this way, we can have a better comparison between different compounds. The recalculated local susceptibilities are shown in Fig.1(d).

{\bf Absolute neutron-scattering intensity normalization}
We performed our inelastic neutron scattering on LiFeAs and LiFe$_{0.88}$Co$_{0.12}$As at the wide Angular-Range Chopper Spectrometer (ARCS) and Cold Neutron Chopper Spectrometer (CNCS) at Spallation Neutron Source, Oak Ridge National Laboratory and on LiFe$_{0.955}$V$_{0.045}$As at Four-Dimensional (4D) Space Access Neutron Spectrometer (4SEASONS), Materials and Life Science Experimental Facility (MLF), Japan Proton Accelerator Research Complex (J-PARC). The neutron scattering intensity of the LiFeAs and LiFe$_{0.88}$Co$_{0.12}$As have already been normalized by standard vanadium sample. The neutron scattering intensity of LiFe$_{0.955}$V$_{0.045}$As is normalized by phonon intensities since a standard Vanadium is unavailable during the experiment. As discussed in the main text, V-normalization is actually less reliable in V-doped LeFeAs because of the unknown amount of V-containing flux in the sample. Therefore, phonon normalization using the same phonon for LiFeAs, which we re-analyzed from using previous data, and V-doped LiFeAs should be more reliable than the standard Vanadium scan when comparing the
results from different spectrometers \cite{Scott2016}.

We identified two phonons, one is around $(1,1)$ and the other originates from $(2,0)$. The incident neutron energy is almost the same, about 39 meV for LiFe$_{0.955}$V$_{0.045}$As and 35 meV for LiFeAs and LiFe$_{0.88}$Co$_{0.12}$As.  The one-dimensional (1D) cuts of these phonons are shown in Fig. 7, and the integrated intensity of each phonon can then be calculated. By comparing the phonon intensities of different materials, we estimated the scale factor which is defined as the ratio of phonon intensity between LiFeAs/LiFe$_{0.88}$Co$_{0.12}$As and LiFe$_{0.955}$V$_{0.045}$As. Since the intensity of LiFeAs and LiFe$_{0.88}$Co$_{0.12}$As has been normalized already by the standard Vanadium scan, we could calculate the absolute value of the intensities in LiFe$_{0.955}$V$_{0.045}$As.  The data of LiFeAs and LiFe$_{0.88}$Co$_{0.12}$As can be used as a cross-check and the absolute value we obtained for LiFe$_{0.955}$V$_{0.045}$As is consistent with these two references.

Fig. 7(a) and Fig. 7(b) show the acoustic phonon near $(1,1)$ at 4.5 and 5.5 meV [Fig. 7(e)], and Fig. 7(c) and (d) display another phonon around $(2,0)$ [Fig. 7(f)] at 12.5 and 13.5 meV.  In Fig. 7(g), we plot the estimated factor from Fig. 7(a-d). An average scale factor about 200 is obtained.

Alternatively, the incoherent nuclear scattering can be used as another reference to normalize the data. In Fig. 8, we show the incoherent peak for LiFe$_{0.955}$V$_{0.045}$As, LiFeAs and LiFe$_{0.88}$Co$_{0.12}$As, respectively. The incident energy $E_i$ is 39 and 75 meV for LiFe$_{0.955}$V$_{0.045}$As and 35 and 80 meV for LiFeAs and LiFe$_{0.88}$Co$_{0.12}$As. The estimated scale factor is about 70, significantly smaller than the value from phonon normalization. We note here that the residual flux trapped in the single crystal samples is an issue that may affect the estimation of the absolute value. This has been observed in our previous paper on Co-doped NaFeAs in which a difference of 30\% was observed \cite{Scott2016}. This situation would be even more serious since the LiFe$_{0.955}$V$_{0.045}$As actually was grown out of a nominal composition of LiFe$_{0.9}$V$_{0.1}$As. In particular, the neutron incoherent scattering cross section of V is huge compared with that of Fe, Li, and As.  The trapped flux with rich V having a large neutron incoherent cross section would lead to a serious underestimation of the scale factor. Therefore, we argue that the phonon normalization is more accurate than the normalization by incoherent peaks.

More importantly, we emphasize that even with this smaller scale ratio, we can still observe the enhancement of magnetic scattering in LiFe$_{0.955}$V$_{0.045}$As [Fig. 9], demonstrating the robustness of our conclusion that the low energy magnetic excitations in V-doped LiFeAs is greatly enhanced.  In the main text, we used the scale factor ~200 to normalize all the data. In Fig.9, we display the local susceptibility of LiFe$_{0.955}$V$_{0.045}$As from both normalization methods, as well as others for comparison.

{\bf Transport results}
Figure 10 shows the magnetic susceptibility and electric resistivity in a series of V-doped LiFeAs. The effective magnetic moments were obtained from the fits with a Curie-Weiss form
${1\over \chi-\chi_0}=(T-\theta)/C$, in which $\theta$ is the Curie temperature and $\chi_0$ is the temperature independent constant. It is clear that the effective magnetic moment increasing with the concentration of Vanadium, as show in inset of Figure 10(a). Figure 10(b) shows the temperature dependence of resistivity of LiFe$_{1-x}$V$_x$As varies from $x=0$ to 0.15, showing the evolution from metallic to insulating behavior. A non-Fermi liquid region was identified with $x$ between 0.05 and 0.15, suggesting enhanced electronic correlation [Fig. 10(b) inset]. We note that although the susceptibility and resistivity probe the energy scales different from neutron scattering, the increase of moments and resistivity is qualitatively consistent with the enhancement of low energy magnetic excitations.

{\bf DFT+DMFT calculations}
We use density functional theory combined with dynamical mean field theory (DFT+DMFT) \cite{Kotliar06} to compute the electronic structure and spin dynamics of V-doped LiFeAs in the paramagnetic state. The density functional theory part is based on the full-potential linear augmented plane wave method implemented in Wien2K \cite{Blaha01} in conjunction with Perdew–Burke–Ernzerhof generalized gradient approximation \cite{Perdew96} of the exchange correlation functional. DFT+DMFT is implemented on top of Wien2K and documented in Ref. \cite{Haule10}. In the DFT+DMFT calculations, the electronic charge was computed self-consistently on DFT+DMFT density matrix. The quantum impurity problem was solved by the continuous time quantum Monte Carlo (CTQMC) method \cite{Haule07,Werner06} with a Hubbard $U=5.0$ eV and Hund's rule coupling $J=0.8$ eV in the paramagnetic state \cite{Yin2011,YinN11,YinN14}. Bethe-Salpeter equation is used to compute the dynamic spin susceptibility where the bare susceptibility is computed using the converged DFT+DMFT Green’s function while the two-particle vertex is directly sampled using CTQMC method after achieving full self-consistency of DFT+DMFT density matrix. The detailed method of computing the dynamic spin susceptibility is documented in Ref. \cite{Yin2011} and was shown to be able to compute accurately the spin dynamics of many iron pnictide superconductors. The experimental crystal structure (space group I4/mmm, \#139) of V-doped LiFeAs with lattice constants $a=b= 3.7914$ \AA\ and $c=6.3639$ \AA\ \cite{Tapp} is used in the calculations. Virtual crystal approximation is employed to approximate the V-doping effect. In Figure 11, we plot the calculated dynamic spin susceptibility from two inter-orbital scattering channels which are dxy-dxz and dxy-dyz in LiFeAs and LiFe$_{0.967}$V$_{0.033}$As. The enhancement of electron scattering in both channels in LiFe$_{0.967}$V$_{0.033}$As is closely associated with the Fermi surface nesting between the inner hole pockets ($d_{xz}/d_{yz}$) and the inner electron pocket ($d_{xy}$). Such an enhanced electron scattering between different orbitals is consistent with the “spin mixing” process as shown in Fig. 4(c) in the manuscript and drives the system into the $S=1/2$ state. Furthermore, such spin-state fluctuations might be responsible for the biquadratic exchange coupling which is essential for the nematic instability existing extensively in iron pnictide/calcogenide superconductors.

{\bf DATA AVAILABILITY}
The neutron scattering and numerical data used in this work, and in the figures of this
manuscript, are available upon request from the corresponding authors.

{\bf CODE AVAILABILITY}
The computer codes used to carry out the DFT+DMFT calculations used in this
work are available upon request from the corresponding author.

{\bf ACKNOWLEDGEMENTS}
The  neutron scattering work at Beijing Normal University is supported by the
Fundamental Research Funds for the Central Universities
(Grants No. 310432101 and No. 2014JJCB27) and the
National Natural Science Foundation of China (Grant No.
11734002). Z.P.Y. was supported by the NSFC (Grant No. 11674030), the Fundamental Research Funds for the Central Universities (Grant No. 310421113), the National Key Research and Development Program of China grant 2016YFA0302300. The calculations used high performance computing clusters at BNU in Zhuhai and the National Supercomputer Center in Guangzhou.
The neutron scattering work at Rice is supported by the U.S. DOE BES under contract no.
DE-SC0012311 (P.D.). The neutron experiment at the Materials and Life Science Experimental Facility of the J-PARC was performed under a user program 2017B0216.

{\bf AUTHOR CONTRIBUTIONS}
P.D. G.T.T. X.L. and C.Q.J. designed the research. Single crystals were grown by G.Y.D. and L.Y.X.,
under the instructions of X.C.W. and C.Q.J.  Neutron scattering experiments were carried out by Z.X., G.T.T., Y.R., L.T., Y.L., P.P.L., H.W., Y.W., R.K., K.I., and D.L.A.  Theoretical calculations were carried out by Z.P.Y.
The paper was written by P.D., Y.L. Z.X., and Z.P.Y. All authors made comments on the paper.

{\bf COMPETING INTERESTS}
The authors declare no competing interests.

{\bf ADDITIONAL INFORMATION}
Supplementary information is available for this paper at

{}

%\counterwithout{figure}{section}
%\captionstyle{normal}
\begin{figure}
\includegraphics[scale=.45]{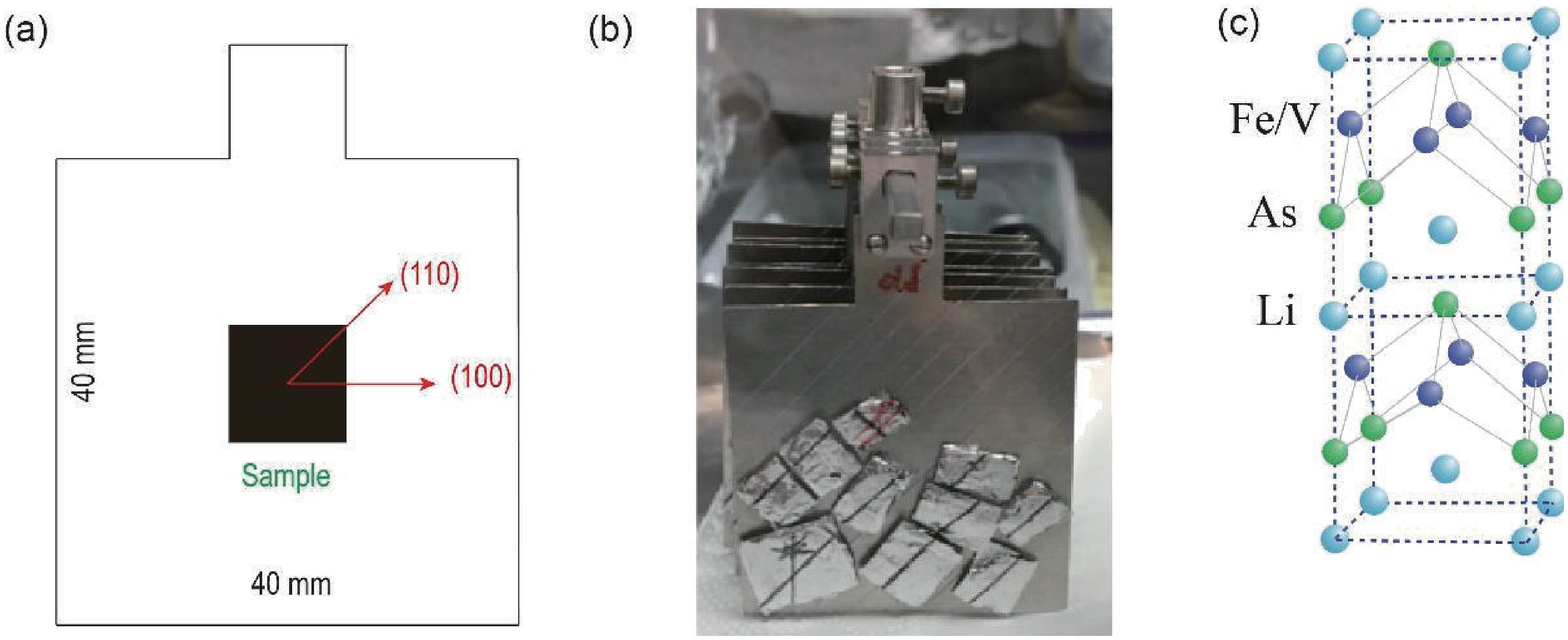}
\caption{(Color online) (a),(b) are the schematic and physical image of the configuration, separately. The [110] direction of all the samples is along the diagonal direction of the Aluminum plane. (c) The schematic chemical structure of LiFe$_{0.955}$V$_{0.045}$As. Two tetragonal unit cells stacked along the c axis are shown.}
\end{figure}

\begin{figure}
	\includegraphics[scale=.45]{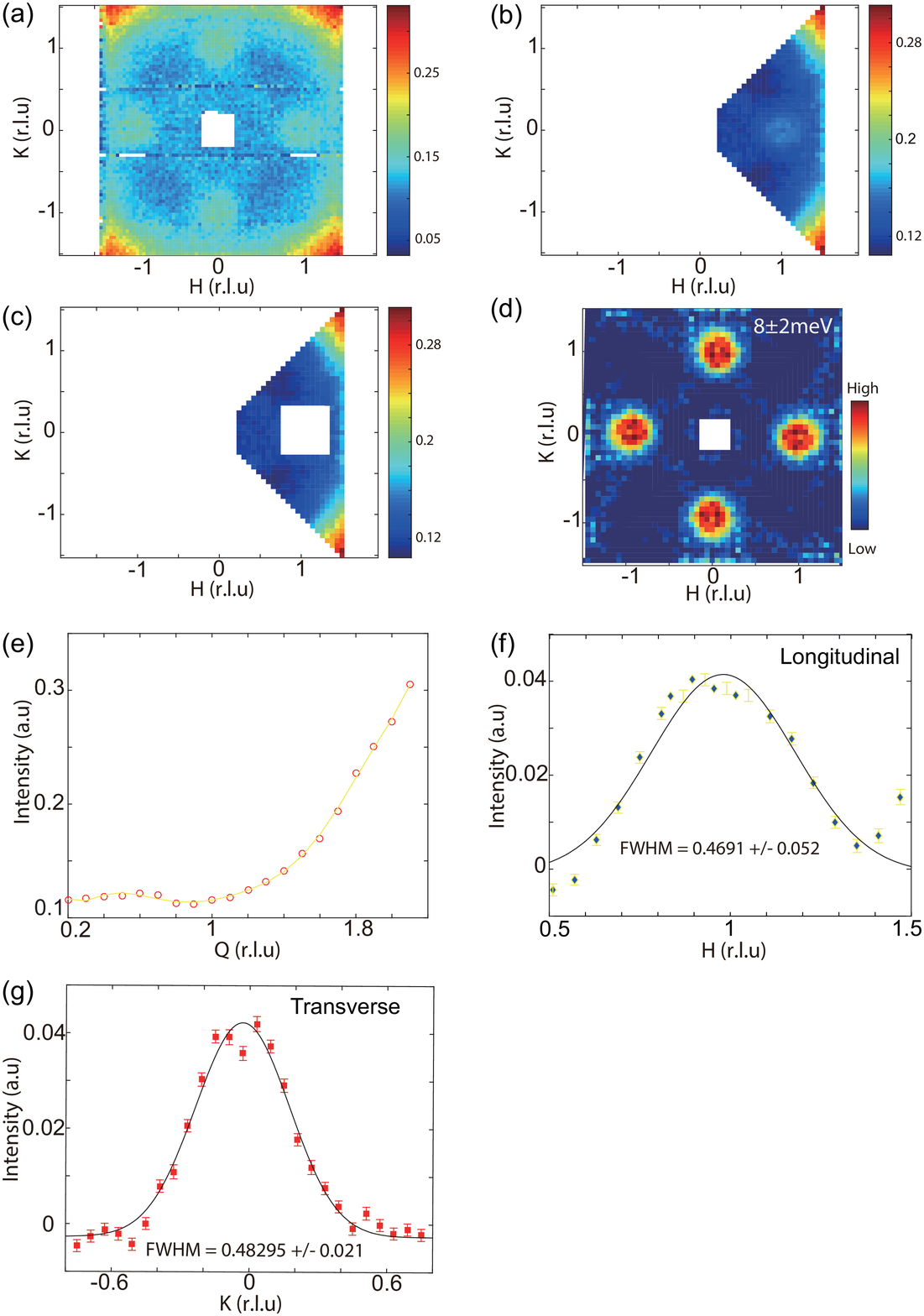}
	\caption{ (Color online) (a). Constant energy map of the raw data at E = [6,10] meV in LiFe$_{0.955}$V$_{0.045}$As. Clear fourfold symmetric magnetic signals are showed. (b) Folded image of frame (a).  (c) The background intensity plot with the magnetic signal masked by the white square. (d) Magnetic signal after background subtraction. (e) Radially symmetric background as a polynomial function of $\left|Q\right|$ from (c). (f),(g) 1-d cut of the spin excitations along the longitudinal direction transverse direction, respectively. The peaks are fitted with Gaussian function.}
\end{figure}

\begin{figure}[!t]
	\includegraphics[scale=.45]{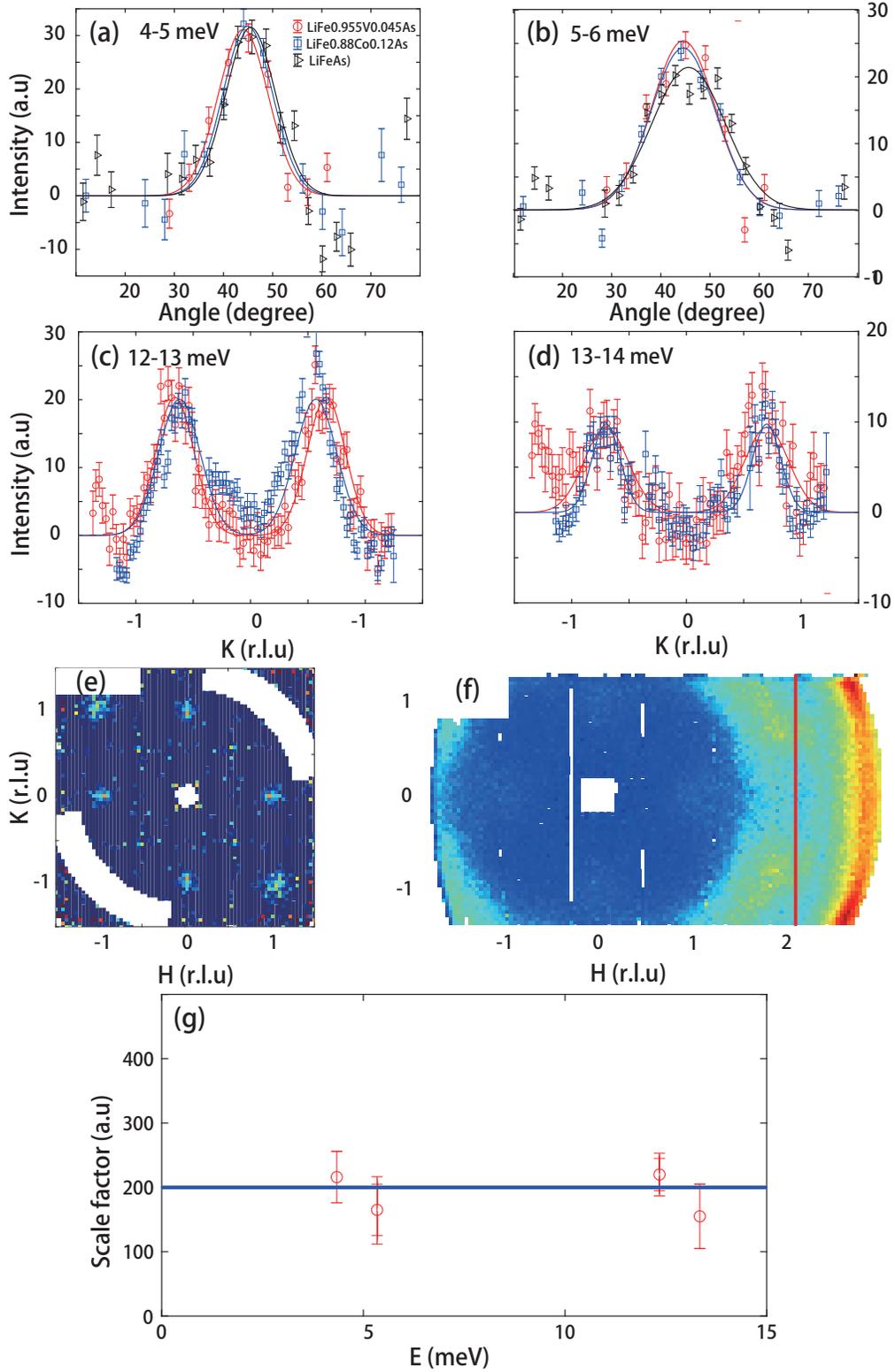}
	\caption{ (Color online) (a-d) One-dimensional cuts of the phonon intensities of LiFeAs, LiFe$_{0.955}$V$_{0.045}$As and LiFe$_{0.88}$Co$_{0.12}$As at energy transfer $4.5\pm 0.5$ meV (a), $5\pm 0.5$ meV (b), $12.5\pm 0.5$ meV (c), $13.5\pm 0.5$ meV (d) respectively. The scale factors have been multiplied to normalize the intensity. (e),(f) Two-dimensional slices of the neutron scattering intensity in LiFe$_{0.88}$Co$_{0.12}$As at 4meV and 13 meV, respectively. The red line represent the cut direction for frame (c) and (d).  (g) The scale factors obtained at several energies.}
\end{figure}

\begin{figure}[!t]
	\includegraphics[scale=.45]{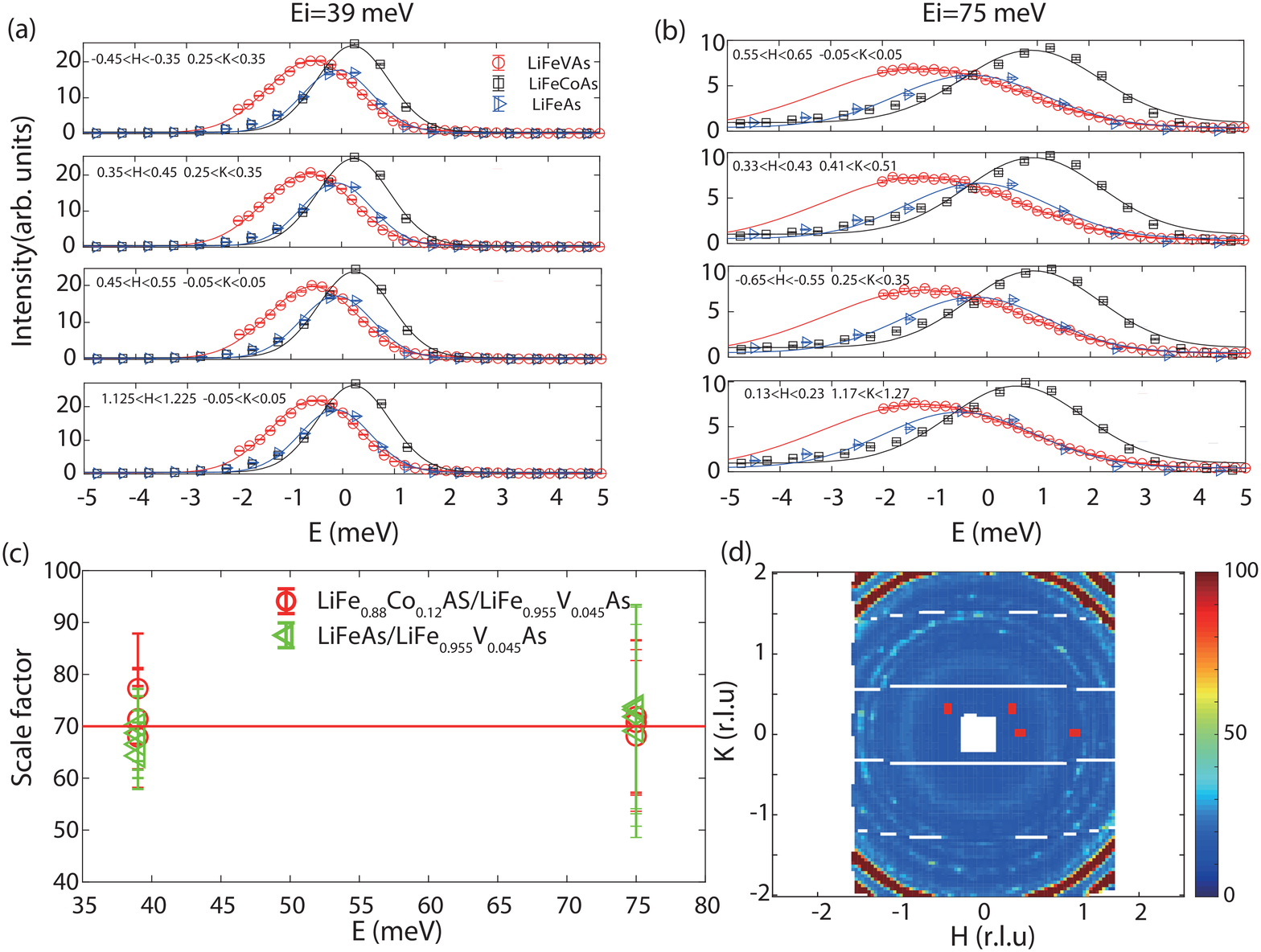}
	\caption{ (Color online) (a),(b) Incoherent nuclear scattering at the elastic line for Ei = 39 meV and 75 meV , respectively.  (c) The scale factor obtained by normalization with incoherent scattering. (d) The location of the four cuts from frame (a) and (b) in H-K plane is marked by red rectangle.}
\end{figure}

\begin{figure}[!t]
	\includegraphics[scale=.45]{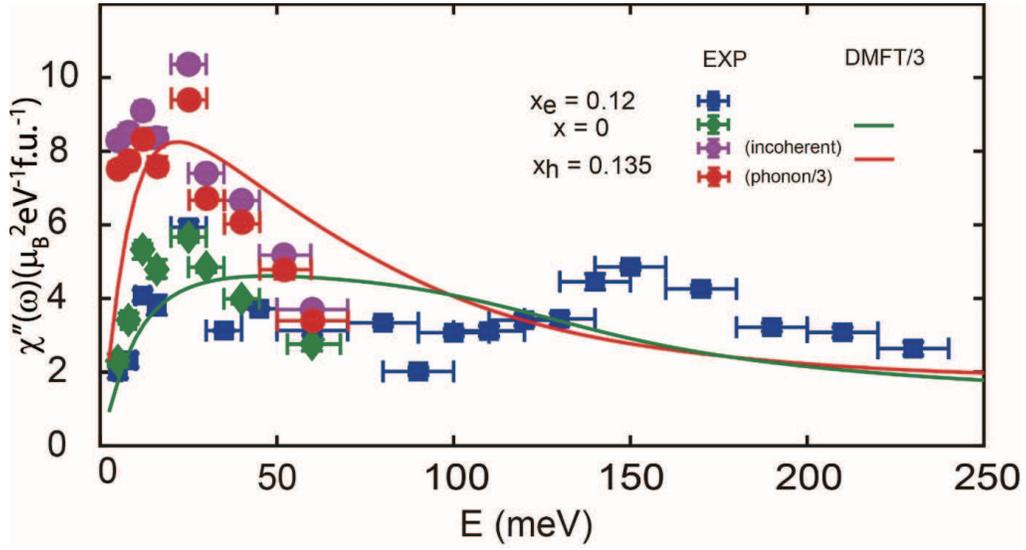}
	\caption{ (Color online) Theoretical and experimental local spin susceptibility in LiFeAs, LiFe$_{0.955}$V$_{0.045}$As and LiFe$_{0.88}$Co$_{0.12}$As. The purple solid circle is the one normalized by incoherent elastic scattering and the red solid circle is normalized by phonon with a 1/3 factor. Green and red solid lines represent the DMFT results divided by a factor of 3.}
\end{figure}

\begin{figure}[!t]
	\includegraphics[scale=.45]{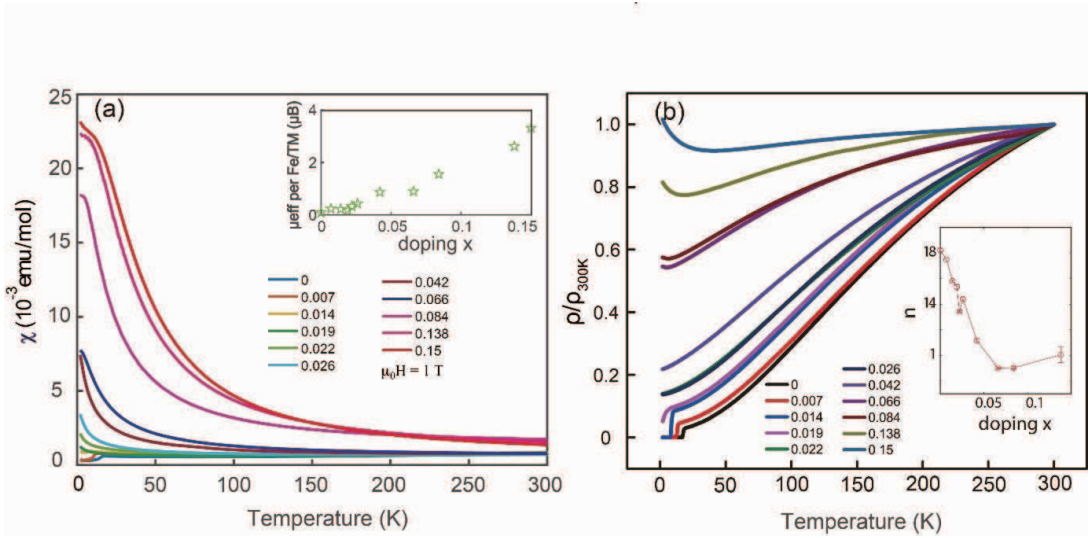}
	\caption{ (Color online) The magnetic susceptibility (a) and normalized electric resistivity${\rho}/{\rho_{300K}}$ (b) in a variety of V-doped LiFeAs. The inset of (a) is the doping dependence of the effective magnetic moment from fits to Curie-Weiss form. The inset of (b) is the doping dependence of the n which is related with $\rho=\rho_0+aT^n$ }
\end{figure}

\begin{figure}[!t]
	\includegraphics[scale=.45]{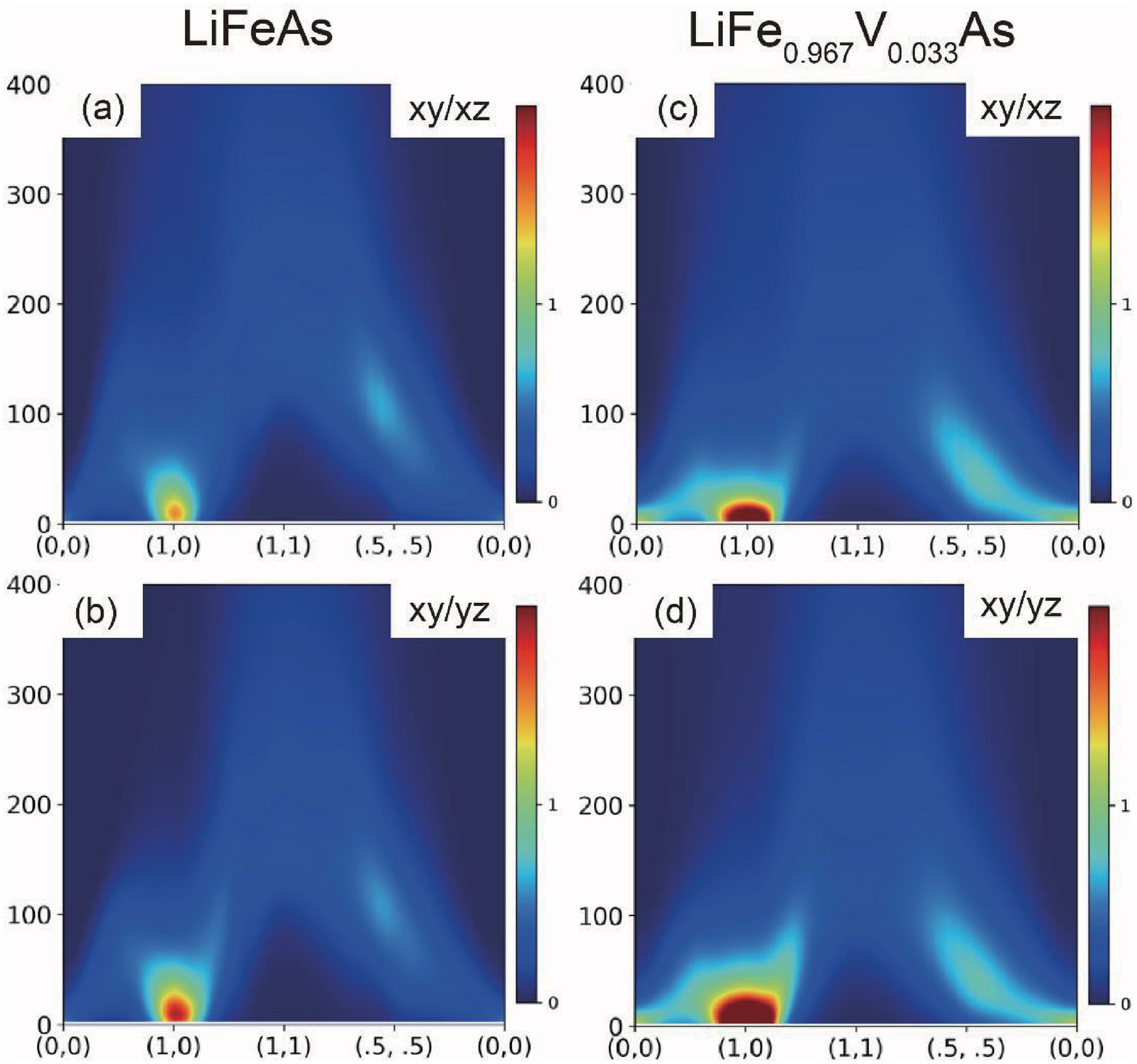}
	\caption{ (color online) Dynamic spin susceptibility from two inter-orbital scattering channels (dxy-dxz and dxy-dyz) by DFT+DMFT calculation. In LiFe0.967V0.033As with better Fermi surface nesting between dxy and dxz/yz orbital, the spin dynamic susceptibility from these two channels are greatly enhanced}
\end{figure}


\begin{thebibliography}{}

\bibitem{Mantle} Lin, J. F. {\it et al.}, Intermediate-spin ferrous iron in lowermost mantle post-perovskite and perovskite. Nat. Geosci., {\bf 1}, 688-691 (2008).

\bibitem{Hemoglobin} Bren, K. L., Eisenberg, R., and Gray, H. B., Discovery of the magnetic behavior of hemoglobin: A beginning of bioinorganic chemistry. Proc. Natl. Acad. Sci. U.S.A., {\bf 112}, 13123-13127 (2015).

\bibitem{stewart} Stewart, G. R., Superconductivity in iron compounds. Rev. Mod. Phys. {\bf 83}, 1589-1652 (2011).
\bibitem{Frank2009} Kr\"{u}ger, F., Kumar, S., Zaanen, J., and van den Brink, J., Spin-orbital frustrations and anomalous metallic state in iron-pnictide superconductors. Phys. Rev. B, {\bf 79}, 054504 (2009).

\bibitem{Dai_RMP} Dai, P. C., Antiferromagnetic order and spin dynamics in iron-based superconductors. Rev. Mod. Phys. 87, 855 (2015).
\bibitem{Yin2011} Yin, Z. P., Haule, K., and Kotliar, G., Kinetic frustration and the nature of the magnetic and paramagnetic states in iron pnictides and iron chalcogenides. Nat. Mater., {\bf 10}, 932 (2011).
\bibitem{norman} Mannella, N. The magnetic moment enigma in Fe-based high temperature superconductors. J. Phys.: Condens. Matter {\bf 26}, 473202 (2014).
\bibitem{Lorenzana2005} Lorenzana, J., Seibold, G., and Coldea, R., Sum rules and missing spectral weight in magnetic neutron scattering in the cuprates. Phys. Rev. B, {\bf 72}, 224511 (2005).
\bibitem{Meng2013} Wang, M. {\it et al.}, Doping dependence of spin excitations and its correlations with high-temperature superconductivity in iron pnictides. Nat. Commun. {\bf 4}, 2874 (2013).
\bibitem{Luo2012} Luo, H. Q. {\it et al.}, Electron doping evolution of the anisotropic spin excitations in BaFe$_{2-x}$Ni$_x$As$_2$. Phys. Rev. B, {\bf 86}, 024508 (2012).
\bibitem{Liu2012} Liu, M. S. {\it et al.}, Nature of magnetic excitations in superconducting BaFe$_{1.9}$Ni$_{0.1}$As$_2$. Nat. Phys. {\bf 8}, 376-381 (2012).
\bibitem{Luo2013} Luo, H. Q. {\it et al.}, Electron doping evolution of the magnetic excitations in BaFe$_{2-x}$Ni$_x$As$_2$. Phys. Rev. B, {\bf 88}, 144516 (2013).
\bibitem{Horigane2016} Horigane, K. {\it et al.}, Spin excitations in hole-overdoped iron-based superconductors. Sci. Rep. {\bf 6}, 33303 (2016).
\bibitem{Scott2016} Carr, S. V. {\it et al.}, Electron doping evolution of the magnetic excitations in NaFe$_{1-x}$Co$_x$As. Phys. Rev. B {\bf 93}, 214506 (2016).
\bibitem{Qureshi2012} Qureshi, N. {\it et al.}, Inelastic neutron-scattering measurements of incommensurate magnetic excitations on superconducting LiFeAs single crystals. Phys. Rev. Lett. {\bf 108}, 117001 (2012).
\bibitem{Qureshi2014} Qureshi, N. {\it et al.}, Fine structure of the incommensurate antiferromagnetic fluctuations in single-crystalline LiFeAs studied by inelastic neutron scattering. Phys. Rev. B {\bf 90}, 144503 (2014).
\bibitem{Wang2012} Wang, M. {\it et al.}, Effect of Li-deficiency impurities on the electron-overdoped LiFeAs superconductor. Phys. Rev. B {\bf 86}, 144511 (2012).

\bibitem{Yu2016} Li, Y. {\it et al.}, Orbital Selective Spin Excitations and their Impact on Superconductivity of LiFe$_{1-x}$Co$_x$As. Phys. Rev. Lett. {\bf 116}, 247001 (2016).

\bibitem{Wang2016} Wang, Q. S. {\it et al.}, Magnetic ground state of FeSe. Nat. Commun. {\bf 7}, 12182 (2016).
\bibitem{Zaliznyak2011} Zaliznyak, I. A. {\it et al.}, Unconventional Temperature Enhanced Magnetism in Fe$_{1.1}$Te. Phys. Rev. Lett. {\bf 107}, 216403 (2011).

\bibitem{hirschfeld} Hirschfeld, P. J., Korshunov, M. M., and Mazin, I. I., Gap symmetry and structure of Fe-based superconductors. Rep. Prog. Phys. {\bf 74}, 124508 (2011).

\bibitem{Medici2014} de'Medici, L., Giovannetti, G., and Capone, M., Selective Mott physics as a key to iron superconductors. Phys. Rev. Lett. {\bf 112}, 177001 (2014)

\bibitem{yu2017} Yu, R. and Si, Q., Orbital-selective Mott phase in multiorbital models for iron pnictides and chalcogenides. Phys. Rev. B {\bf 96}, 125110 (2017).

\bibitem{komijani2017} Komijani, Y. and Kotliar, G. Analytical slave-spin mean-field approach to orbital selective Mott insulators. Phys. Rev. B {\bf 96}, 125111 (2017).

\bibitem{Mn1} Gastiasora, M. N. and Anderson, B. M., Enhancement of magnetic stripe order in iron-pnictide superconductors from the interaction between conduction electrons and magnetic impurities. Phys. Rev. Lett. {\bf 113}, 067002 (2014).

\bibitem{Mn2} Inosov, D. S. {\it et al.}, Possible realization of an antiferromagnetic Griffiths phase in Ba(Fe$_{1-x}$Mn$_x$)$_2$As$_2$. Phys. Rev. B {\bf 87}, 224425 (2013).

\bibitem{Tucker2012} Tucker, G. S. {\it et al.}, Competition between stripe and checkerboard magnetic instabilities in Mn-doped BaFe$_2$As$_2$. Phys. Rev. B, {\bf 86}, 020503 (2012).

\bibitem{Song2016} Song, Y. {\it et al.}, A Mott insulator continuously connected to iron pnictide superconductors. Nat. Comm. {\bf 7}, 13879 (2016).

\bibitem{Matt2016} Matt, C. E. {\it et al.}, NaFe$_{0.56}$Cu$_{0.44}$As: A Pnictide Insulating Phase Induced by On-Site Coulomb Interaction. Phys. Rev. Lett. {\bf 117}, 097001 (2016).

\bibitem{Xing2016} Xing, L. Y. {\it et al.}, Observation of non-Fermi liquid behavior in hole-doped LiFe$_{1-x}$V$_x$As. Phys. Rev. B {\bf 94}, 094524 (2016).

\bibitem{Ong2012} Ong, T. T. and Coleman, P. Local quantum criticality of an iron-pnictide tetrahedron. Phys. Rev. Lett. {\bf 108},  107201, (2012).

\bibitem{JHZhang2000} Zhang, J. H., Sknepnek, R., and Schmalian, J., Spectral analysis for the iron-based superconductors: Anisotropic spin fluctuations and fully gapped s$^\pm$-wave superconductivity. Phys. Rev. B {\bf 82}, 134527 (2010).

\bibitem{TChen2019} Chen, T. {\it et al.}, Anisotropic spin fluctuations in detwinned FeSe. Nature Materials {\bf 18}, 709 (2019).

\bibitem{Pfau} Pfau, H. {\it et al.}, Momentum Dependence of the Nematic Order Parameter in Iron-Based Superconductors. Phys. Rev. Lett. {\bf 123}, 066402 (2019).

\bibitem{Pfau2019} Pfau, H. {\it et al.}, Detailed band structure of twinned and detwinned BaFe$_2$As$_2$ studied with angle-resolved photoemission spectroscopy. Phys. Rev. B {\bf 99}, 035118 (2019).

\bibitem{Watson2019} Watson, M. D. {\it et al.}, Probing the reconstructed Fermi surface of antiferromagnetic BaFe$_2$As$_2$ in one domain. npj Quantum Materials {\bf 4}, 36 (2019).

\bibitem{LTian2019} Tian, L. {\it et al.}, Spin fluctuation anisotropy as a probe of orbital-selective hole-electron quasiparticle excitations in detwinned Ba(Fe$_{1-x}$Co$_x$)As$_2$.
Phys. Rev. B {\bf 100}, 134509 (2019).

\bibitem{CZhang2011} Zhang, C. L. {\it et al.}, Neutron Scattering Studies of spin excitations in hole-doped Ba$_{0.67}$K$_{0.33}$Fe$_2$As$_2$ superconductor. Scientific Report {\bf 1}, 115 (2011).

\bibitem{Harriger} Harriger, L. W. {\it et al.}, Nematic spin fluid in the tetragonal phase of BaFe$_2$As$_2$. Phys. Rev. B {\bf 84}, 054544 (2011)

\bibitem{xylu18} Lu, X. Y. {\it et al.}, Spin Waves in Detwinned BaFe$_2$As$_2$. Phys. Rev. Lett. {\bf 121}, 067002 (2018).

\bibitem{Rev_hund} Georges, A., de'Medici, L., and Mravlje, J.,  Strong electronic correlations from Hund's coupling. Annu. Rev. Condens. Matter Phys., {\bf 4}, 137-178 (2013).

\bibitem{Aron2015} Aron, C. and Kotliar, G., Analytic theory of Hund's metals: A renormalization group perspective. Phys. Rev. B {\bf 91},  041110 (2015).

\bibitem{Stadler2015} Stadler, K. M., Yin, Z. P., von Delft, J., Kotliar, G., and Weichselbaum, A., Dynamical mean-field theory plus numerical renormalization-group study of spin-orbital separation in a three-band hund metal. Phys. Rev. Lett. {\bf 115},  136401 (2015).

\bibitem{Chaloupka2013} Chaloupka, J. and Khaliullin, G., Spin-State Crossover Model for the Magnetism of Iron Pnictides. Phys. Rev. Lett. {\bf 110},  207205 (2013).

\bibitem{SCO} Spin Crossover in Transition Metal Compounds I, edited by G\"{u}tlich, P. and
Goodwin, H. A., (Springer, Berlin, 2004).

\bibitem{Gretarsson2013} Gretarsson, H. {\it et al.}, Spin-state transition in the Fe pnictides. Phys. Rev. Lett. {\bf 110}, 047003 (2013)

\bibitem{Klingeler2010} Klingeler, R. {\it et al.}, Local antiferromagnetic correlations in the iron pnictide superconductors LaFeAsO$_{1-x}$F$_x$ and Ca(Fe$_{1-x}$Co$_x$)$_2$As$_2$ as seen via normal-state susceptibility. Phys. Rev. B {\bf 81}, 024506 (2010).

\bibitem{Run2017} Yang, R. {\it et al.}, Observation of an emergent coherent state in the iron-based superconductor KFe$_2$As$_2$. Phys. Rev. B {\bf 96}, 201108 (2017).

\bibitem{Wiecki2018} Wiecki, P. {\it et al.}, Pressure dependence of coherence-incoherence crossover behavior in KFe$_2$As$_2$ observed by resistivity and $^{75}$As-NMR/NQR. Phys. Rev. B {\bf 97}, 064509 (2018).

\bibitem{Hardy2013} Hardy, F. {\it et al.}, Evidence of Strong Correlations and Coherence-Incoherence Crossover in the Iron Pnictide Superconductor KFe$_2$As$_2$. Phys. Rev. Lett. {\bf 111}, 027002 (2013).

\bibitem{Miao2016} Miao, H. {\it et al.}, Orbital-differentiated coherence-incoherence crossover identified by photoemission spectroscopy in LiFeAs. Phys. Rev. B {\bf 94}, 201109 (2016).

\bibitem{Lee2006} Lee, P. A., Nagaosa, N., and Wen, X. G., Doping a Mott insulator: Physics of high-temperature superconductivity. Rev. Mod. Phys. {\bf 78}, 17 (2006).

\bibitem{WYWang} Wang, W. Y. {\it et al.}, Orbital selective neutron spin resonance in underdoped superconducting NaFe$_{0.985}$As$_{0.015}$. Phys. Rev. B {\bf 95}, 094519 (2017).

\bibitem{Meng2011} Wang, M. {\it et al.}, Antiferromagnetic spin excitations in single crystals of nonsuperconducting Li$_{1-x}$FeAs. Phys. Rev. B, {\bf 83}, 220515 (2011).

\bibitem{Xing2014} Xing, L. Y. {\it et al.}, The anomaly Cu doping effects on LiFeAs superconductors. J. Phys.: Condens. Matter {\bf 26}, 435703 (2014).

\bibitem{Kajimoto} Kajimoto, R. {\it et al.}, The Fermi Chopper Spectrometer 4SEASONS at J-PARC. J. Phys. Soc. Jpn. {\bf 80}, SB025 (2011).


\bibitem{Kotliar06} Kotliar, G. {\it et al.}, Electronic structure calculations with dynamical mean-field theory. Rev. Mod. Phys. {\bf 78}, 865 (2006).

\bibitem{Blaha01} Blaha, P., Schwarz, K., Madsen, G., Kvasnicka, D., and Luitz, J., WIEN2K, An Augmented Plane Wave+Local Orbitals Program for Calculating Crystal Properties (Karlheinz Schwarz, Techn. Universit$\rm \ddot{a}$t Wien, Austria, 2001).

\bibitem{Perdew96} Perdew, J. P., Burke, K., and Ernzerhof, M., Generalized Gradient Approximation Made Simple. Phys. Rev. Lett. {\bf 77}, 3865 (1996).

\bibitem{Haule10} Haule, K., Yee, C.-H., and Kim, K., Dynamical mean-field theory within the full-potential methods: Electronic structure of CeIrIn$_4$, CeCoIn$_5$, and CeRhIn$_5$. Phys. Rev. B {\bf 81}, 195107 (2010).

\bibitem{Haule07} Haule, K., Phys. Quantum Monte Carlo impurity solver for cluster dynamical mean-field theory and electronic structure calculations with adjustable cluster base. Rev. B {\bf 75}, 155113 (2007).

\bibitem{Werner06} Werner, P.,  Comanac, A., de'Medici, L., Troyer, M., and Millis, A. J., Continuous-time solver for quantum impurity models.
Phys. Rev. Lett. {\bf 97}, 076405 (2006).

\bibitem{YinN11} Yin, Z. P., Haule, K., and Kotliar, G., Magnetism and charge dynamics in iron pnictides. Nat. Phys. {\bf 7}, 294 (2011).

\bibitem{YinN14} Yin, Z. P., Haule, K., and Kotliar, G., Spin dynamics and orbital-antiphase pairing symmetry in iron-based superconductors. Nat. Phys. {\bf 10}, 845 (2014).

\bibitem{Tapp} Tapp, J. H. {\it et al.}, LiFeAs: An intrinsic FeAs-based superconductor with $T_c=18$ K. Phys. Rev. B {\bf 78}, 060505(R) (2008).

\end{thebibliography}
\end{document}